# MetaCQ: An etextbook platform with an Open Learner Model to support Metacognition.

Beier Wang[1] and Xueting Huang[2]

[1] University of Sydney, Sydney, New South Wales, Australia
[2] Swinburne University of Technology, Melbourne, Victoria, Australia
xuetinghuang@swin.edu.au

**Abstract.** This study has proposed an E-textbook platform, MetaCQ, which integrates ITS and OLM to enable users to monitor their study progress. The platform adopts a chatbot to generate MCQs and man- age learners' study data and their learning model. Additionally, it regulates help-seeking behaviour and provides immediate feedback tailored to users' learning processes. Three adaptive feedback methods have been implemented to construct chatbots, examining the MCQs' relevancy and difficulty through the ThinkAloud study to evaluate the most effective method of measuring the user's study performance. However, no valid result demonstrates which method can significantly assess learners' study outcomes based on the current experiment, which requires further studies to improve it.

## 1 Introduction

Since the development of artificial intelligence and its extension of application in the educational field, along with the shifting study mode into online study due to the pandemic, E-textbooks have become an essential component within the educational technology field from their accessibility and adaptability to assist students' learning process and teachers' teaching procedure being personalisation and convenience.

Intelligent Tutoring System (ITS) is a core feature to support E-textbook functionalities, which has been defined by Nwana[9] as made up of computer programs to support the tutoring process by integrating the teaching materials and guiding the teaching instruction, composing with expert knowledge module, student model module, tutoring module and user interface module. Bull[2] has also suggested that embedding ITS with the E-textbook platform enables the system to adapt learner interaction to educational needs and support learners' study outcomes, and can further result in the composition of the Open Learner Model (OLM) when the learners' learning progresses are visible and accessible since OLM is a learner model that is externalised in a form that users can interpret.

Metacognition has been recognised as "higher-order cognition about cognition" stated from Veenman et al.[13], deploying the OLM into the E-textbook platform will facilitate learners' metacognition activity by supporting long-term learning outcomes mentioned by Bull and Kay[3], and reversely providing immediate metacognitive feedback through the OLMs can improve learners' metacognition skills and benefit with their learning process indicated by Roll et al.[10]. Technology's improvement has further accelerated artificial intelligence's implementation. Chatbot, as a significant product, has been identified as a bridge to connect technology and education, contributing to learners' personalised learning and interactive learning experience with its capability, as stated by Clarizia et al.[6]. Applying the chatbot to generate multiple choice questions (MCQs) can accurately measure users' study outcomes, covering a wide knowledge range with broad ranges of difficulty, while enhancing their metacognitive ability as suggested by Simkin and Kuechler[11].

Under the background driven and benefits of the E-textbook platform, this paper introduces an E-textbook platform, MetaCQ, consisting of the ITS and consolidating the OLM and the metacognition strategy, allowing the user to interact with the chatbot and answering the practice questions in the form of MCQs to improve study outcome.

## 2 Design and Implementations

### 2.1 Design Architecture

MetaCQ's architecture design references PUMPT's structure[7], deployed Sphinx and reStructured Text to build the E-Textbook platform, and uses HTML and JavaScript to support its functions. Similar to PUMPT, MetaCQ also runs locally to support the web page. However, instead of collecting the learning data from the local PC and updating it to the cloud storage, it communicates to the chatbot to allow it to handle the learning data and enable it to respond to the download request to update the OLM of the current topic knowledge mastery level. Content and the code have been uploaded on GitHub.

The MetaCQ consists of three components: **Chatbot**, **OLM** and **Metacognition strategy**. According to the ITS framework, MetaCQ, as the platform, handles the function of the user interface module to provide the interface and interactions; OLM, in the role of student model, maintains students' learning progress and outcomes; chatbot delivers the generated MCQs to represent the expert module, cooperating with metacognition strategy with guidance to serve tutoring module. The components within MetaCQ corresponding to the ITS framework have been displayed in Fig. 1.

### 2.2 Implementations

MetaCQ was designed to instruct users' general data privacy knowledge under the content and developed from the University of Sydney's unit, **DATA3406:**

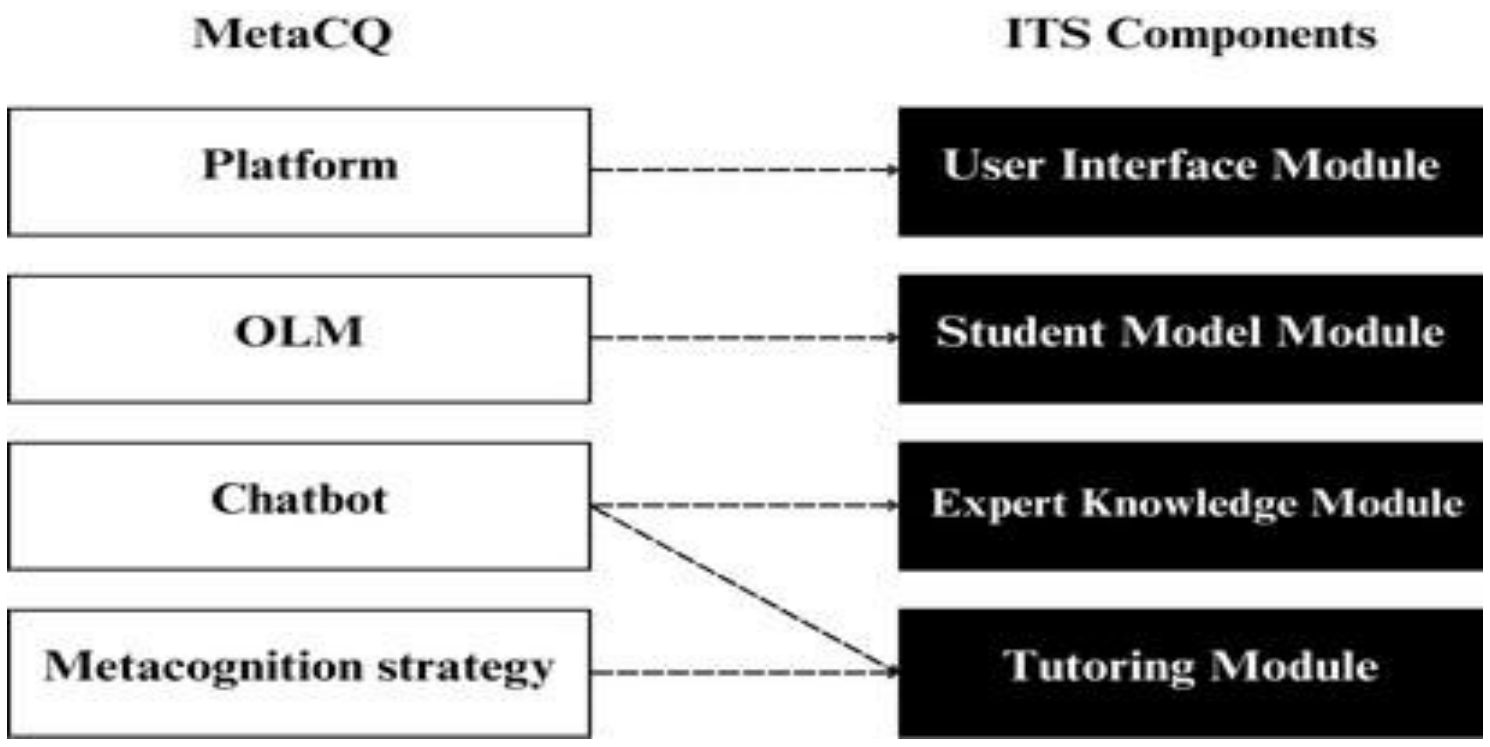

**Fig. 1.** The MetaCQ components and the ITS framework

**Human-in-the-Loop Data Analytics**, which covered the topics of Data privacy principles, Data privacy laws and Data privacy management. Each topic contains clear and corresponding learning objectives for users to reach study expectations and guide the content writing by applying relevant resources to explain theories with simple explanations. To deploy the MetaCQ, users need to read through the content of each topic, answer the MCQs generated from the chatbot and award a satisfied mark to process the next topic's study.

**Chatbot** MetaCQ has applied the chatbot innovated by the University of Sydney, which deployed the GPT-4 large language model to operate the MCQs generation and user's study performance evaluation. Cogniti.ai allows users to create chatbot-based agents that can be steered with specific instructions and resourced with specific contextual information from units of study, as Liu[8] explained. Implementing the on-time chatbot has constituted ELM-ART's limitation of delayed discussion responses[14], which are available with instant interactions to boost user engagement.

MetaCQ has developed three different adaptive feedback method chatbots based on the number of chapter topics. Each topic has assigned a unique adaptive method chatbot to generate the content-aligned questions, minimising each chatbot's content-loading, focusing on the current topic to ensure the question's content quality, and avoiding questions accidentally encountered with other topics. Each practice session offers the user five questions in total, with two marks worth each question. Final scores correspond to different knowledge mastery rankings, as shown in Fig. 2.

Considering the purpose of allowing chatbots to generate effective MCQs, particularly from Butler's six requirements[5] of designing MCQ, chatbots' configuration regarding generating MCQs has focused on: (1) Provide three potential answers to balance the difficulty level with the ideally intermediate level.
(2) The evaluation will deliver concise feedback on each question and overall performance. (3) Question options will avoid the terms of "None-of-the-Above" and "All-of-the-Above". (4) Implement the metacognition strategy, help-seeking, to allow users to reflect in depth on their knowledge.

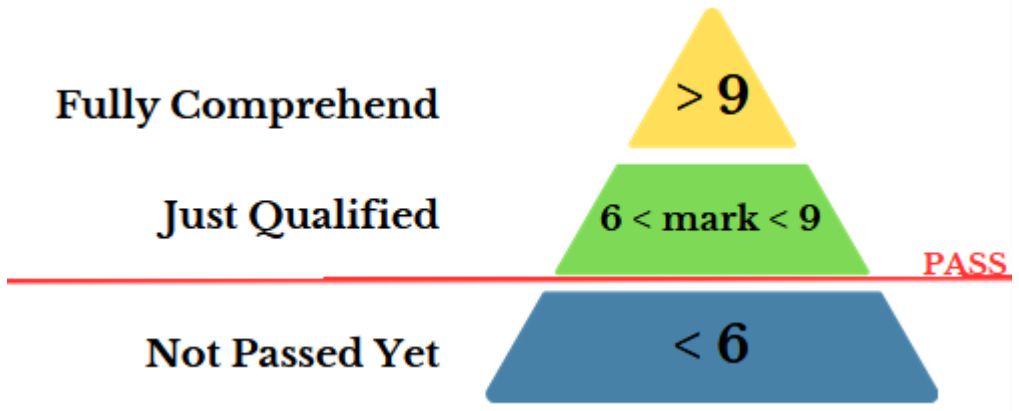

**Fig. 2.** Final mark and knowledge mastery level ranking

To maintain the intermediate level of the MCQ difficulty level, motivated by ELM-ART's weighting strategy[14], each question's default difficulty level has been set at 0.5, ranging from 0 (minimum) to 1 (maximum). Three adaptive feedback methods have been differentiated from the approach of calculating the practice questions' difficulty level: (1) **all-in-all:** The newly generated MCQs practice difficulty level will be based on the previous performance. (2) **1-after-1:** Each newly generated MCQ will be based on the previous MCQ's performance.
(3) N**o extra settings.** The mechanism behind the adaptive feedback methods can be refined into the Fig. 3.

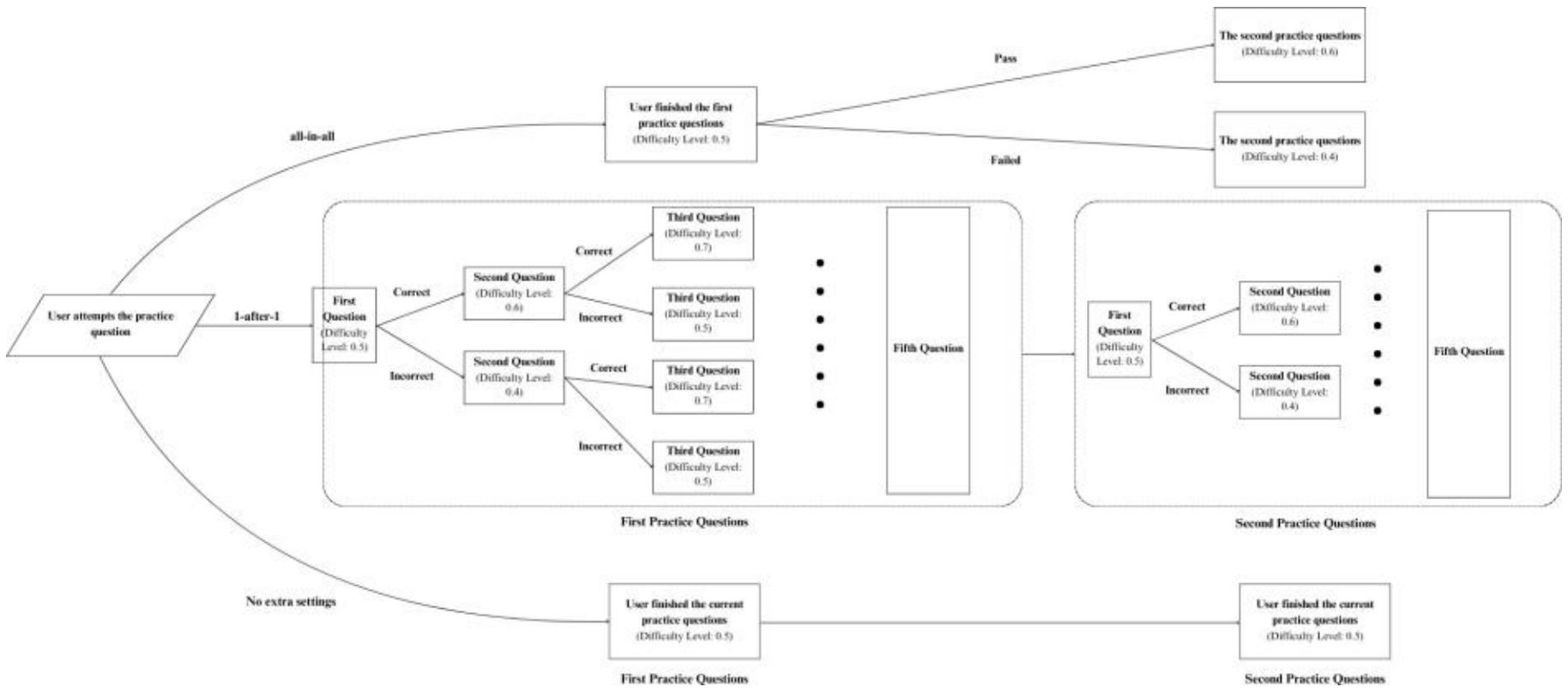

**Fig. 3.** Three adapative feedback methods

**OLM** Within the MetaCQ, there is a table on the homepage that displays the user's current study performance in Fig. 4. All the chapters' knowledge-mastery levels have been initially set as "Not Qualified" to stimulate users to proceed with their study progress and enable users to update their learner models regarding ELM-ART's editable learner model [14]. To update the OLM knowledge mastery levels, users must manually download the practice test conversation between them and ChatGPT and upload the conversation file to the current chapter performance box locally. Users can intuitively visualise their knowledge mastery levels at each topic, be aware of their knowledge mastery status and adjust their study plans. Users can also make reasonable arguments to request the chatbot re-evaluate their knowledge mastery levels by requesting to redo the practice test within the chat. This negotiation process has referred to Bull's persuadable model in the SMIL framework[4], but they do not have access to modify their learner models directly.

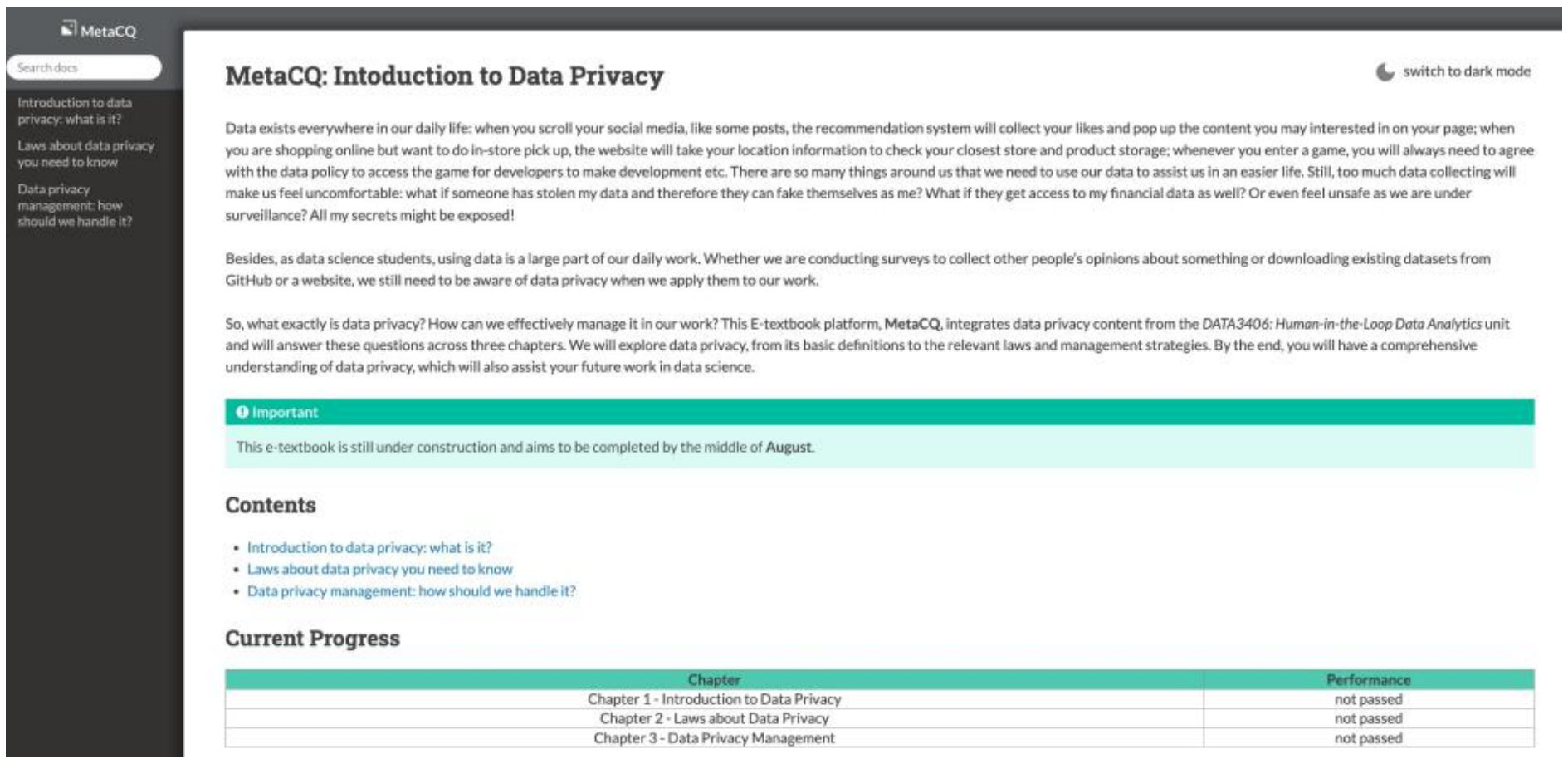

**Fig. 4.** MetaCQ Homepage with OLM

**Metacognition** Research has shown that immediate metacognitive feedback through OLM enhances users' metacognitive skills and supports learning[1], leading to its integration into E-textbook platforms. Building on this, MetaCQ incorporates a help-seeking strategy inspired by Help Tutor. It combines with the question-answering process to provide hints while applying a proportional penalty to marks for assistance. Users can seek help anytime during the answering process by typing "?tips" illustrated as Fig. 5.

The restriction boundary has been claimed to minimise help-seeking errors: users can seek help up to three times per question, with each request revealing additional hints and reducing the attainable marks accordingly. The help-seeking mechanism is illustrated in Fig. 6. Limiting help-seeking opportunities minimises the risk of over-reliance on hints while encouraging users to reflect on knowledge gaps, remain aware of mark deductions, and make informed decisions.

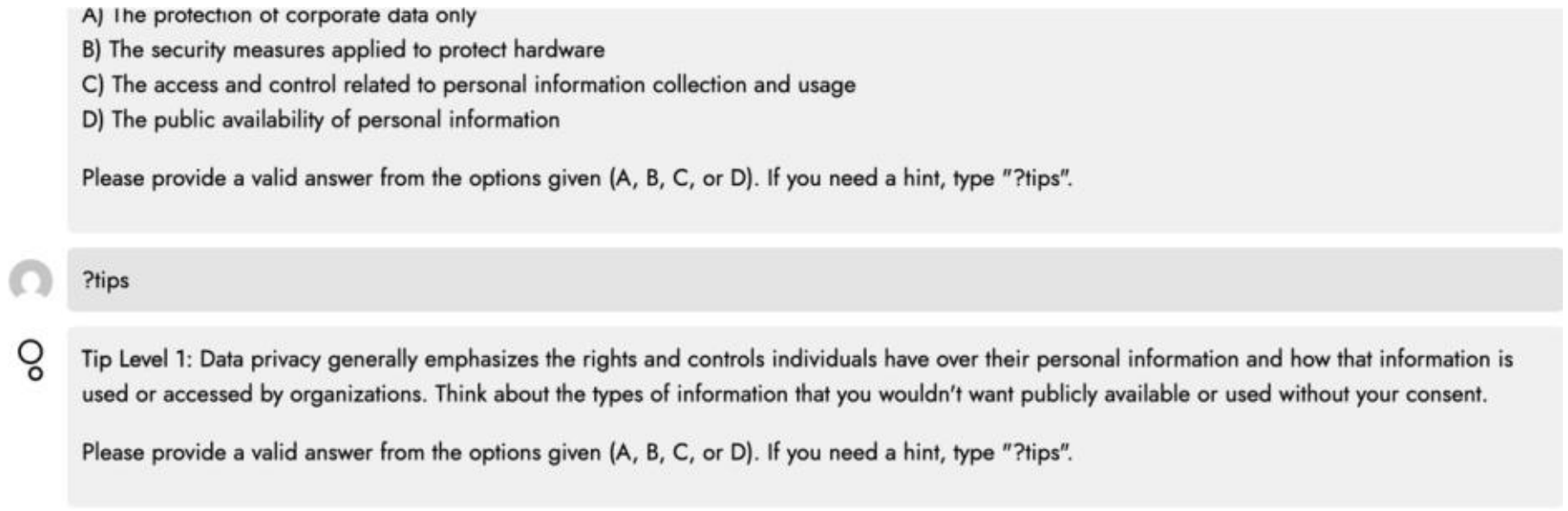

**Fig. 5.** Applying hints during the answering process

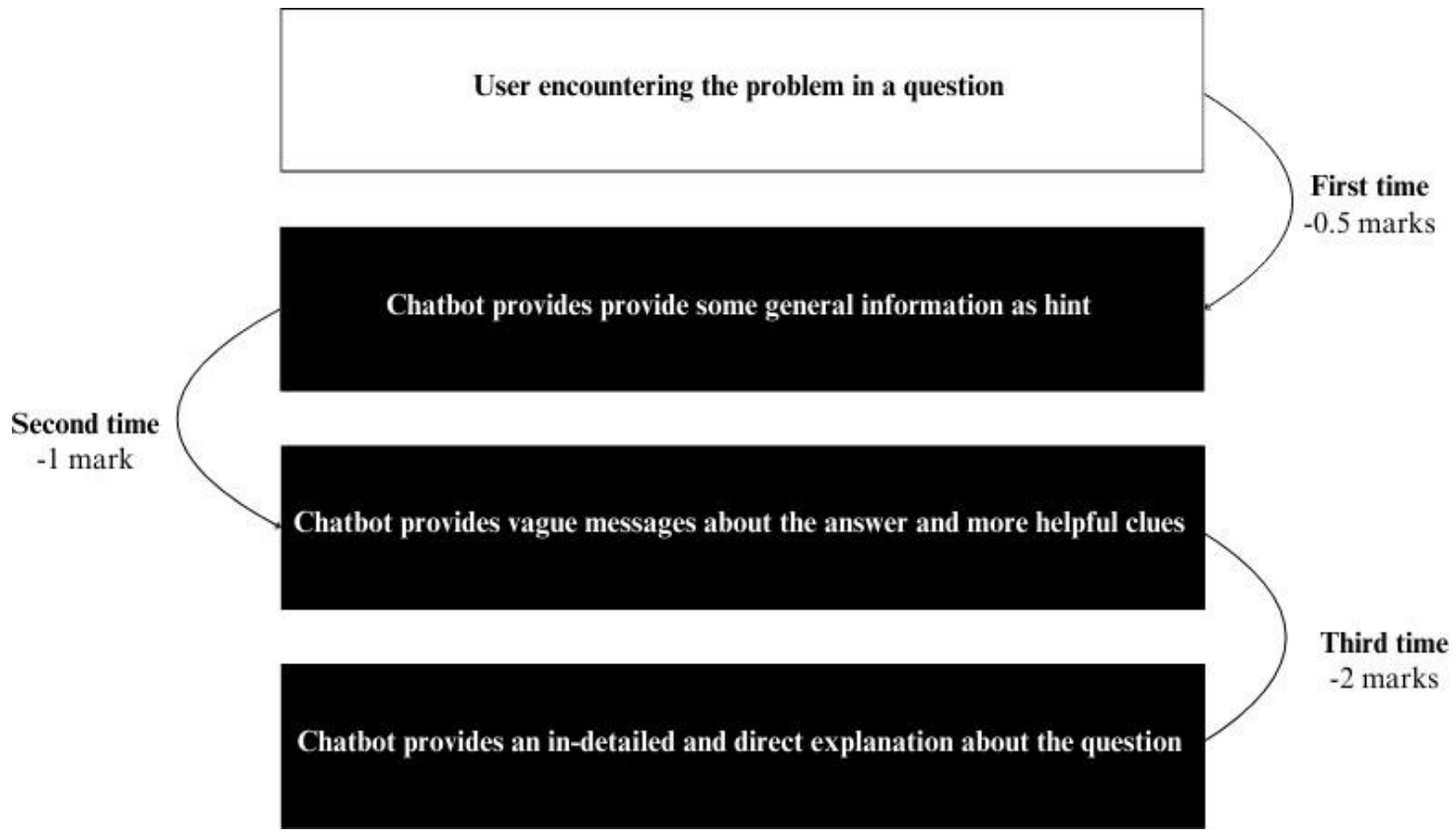

**Fig. 6.** Help-seeking mechanism

## 3 User Study

To examine which adaptive feedback method can more effectively evaluate a user's study performance, a ThinkAloud session was constructed since considering the difficulty level can appropriate tailoring the question based on the user's current knowledge level and enabling the user's learning motivation. Addition- ally, the dynamic difficulty level mechanism that generates MCQs with varied difficulty level points from the chatbot's prompt can further assess the user's depth of understanding.

ThinkAloud is a technique that collects participants' perspectives from verbal communication through tasks that align with the research purpose. Researches [12] have summarised the ThinkAloud flow that the subject needs to talk aloud while solving a problem, and this request is repeated if necessary during the problem-solving process. This procedure encourages participants to tell what they are thinking with veridical reflections of cognitions, allowing researchers to comprehend participants' thoughts, catch occurring problems, and detect minor potential problems. Thus, participants can convey their insights towards each chatbot-generated MCQ, provide accurate and intuitive comments regarding the study tasks, and investigate which prompt method is the most effective way to evaluate a user's learning outcomes.

The ThinkAloud session has been conceptualised into four parts. The "Introduction" aims to guide the participants with the study purpose and grant them the persona of "the student who is currently taking DATA3406 unit's study" and provides a short demonstration to display MetaCQ's workflow in "Demonstration". "Tasks" outlined what participants need to accomplish, and the "Demographic" gathered participants' background knowledge and DATA3406 study status.

Ten participants from the University of Sydney have been selected to support the study's validity, regarding to their relevant backgrounds and data privacy understandings, particularly those participants have accomplished or are currently undertaking the DATA3406 unit.

| Chapter 1 | Chapter 2 | Chapter 3 |
|---|---|---|
| 1-after-1 | No extra settings | all-in-all |

**Table 1.** Chatbot allocations on each chapter

There are two scenarios within the tasks: (1) a student currently working on this material to study data privacy-related knowledge; (2) preparing for the up- coming quiz and reviewing this material as revision. For each task, participants need to evaluate each MCQ's difficulty level after they finish the current practice questions for each chapter, with a Likert scale measurement from 1 to 5, and then explain the rationale behind it. 1 represents very easy, and 5 means very hard. Each chatbot's allocation to each chapter remains unrevealed to reduce the bias within the feedback. The chatbot allocations are demonstrated in Table 1. The repeated tasks allow participants to reach out to more generated MCQs and provide more extensive information to evaluate the generated MCQs' qualities under the different knowledge familiarity scenarios. It also ensures fairness among the adaptive methods examination as "all-in-all" requires the previous test results. During the ThinkAloud session, the conversation pattern has been adhered to as below:

*Host: How would you rate this question's difficulty level with the scale from 1 to 5? While 1 stands for . . .*
*Participant: I will rate this question as . . .*
*Host: Would you like to explain why you would rate this question as . . . ?*
*Participant: I think . . .*

## 4 Results and Discussion

The difficulty level results have been retrieved from participants' Likert scales and their clarified rating reasons. This scale can intuitively measure participants' degrees of opinion and offers comprehensive insight with flexibility from their rationale in the analysis of the results.

Fig. 7 has recorded each participant's difficulty score for each question in two tasks, while Table 2 and Table 3 have calculated each chatbot's difficulty ratings among all the participants with mean, median, and mode scores for each task.

| | | User 1 | User 2 | User 3 | User 4 | User 5 | User 6 | User 7 | User 8 | User 9 | User 10 |
|---|---|---|---|---|---|---|---|---|---|---|---|
| | First Task | | | | | | | | | | |
| | C1Q1 | 1 | 1 | 4 | 2 | 1 | 1 | 1 | 2 | 1 | 1 |
| | C1Q2 | 1 | 1 | 4 | 2 | 2 | 1 | 4 | 2 | 1 | 1 |
| | C1Q3 | 1 | 1 | 3 | 1 | 1 | 1 | 2 | 1 | 1 | 3 |
| | C1Q4 | 5 | 3 | 2 | 2 | 1 | 1 | 1 | 5 | 3 | 3 |
| **1-after-1** | C1Q5 | 4 | 1 | 2 | 1 | 4 | 4 | 1 | 2 | 2 | 1 |
| | C2Q1 | 1 | 2 | 3 | 4 | 2 | 4 | 3 | 1 | 1 | 3 |
| | C2Q2 | 4 | 2 | 2 | 3 | 1 | 2 | 1 | 3 | 1 | 4 |
| | C2Q3 | 1 | 2 | 3 | 1 | 2 | 1 | 3 | 2 | 2 | 1 |
| | C2Q4 | 1 | 1 | 2 | 1 | 2 | 2 | 2 | 1 | 1 | 3 |
| **No Extra Settings** | C2Q5 | 1 | 1 | 2 | 1 | 3 | 1 | 2 | 2 | 1 | 3 |
| | C3Q1 | 1 | 1 | 4 | 1 | 1 | 5 | 1 | 2 | 4 | 1 |
| | C3Q2 | 2 | 1 | 5 | 1 | 2 | 1 | 1 | 4 | 2 | 1 |
| | C3Q3 | 1 | 1 | 1 | 1 | 3 | 1 | 2 | 1 | 1 | 4 |
| | C3Q4 | 1 | 1 | 3 | 2 | 3 | 3 | 4 | 2 | 4 | 1 |
| **all-in-all** | C3Q5 | 4 | 1 | 4 | 1 | 3 | 1 | 1 | 3 | 1 | 4 |
| | SecondTask | | | | | | | | | | |
| | C1Q1 | 1 | 1 | 2 | 1 | 2 | 2 | 2 | 1 | 1 | 1 |
| | C1Q2 | 1 | 1 | 1 | 1 | 2 | 2 | 1 | 1 | 2 | 1 |
| | C1Q3 | 1 | 1 | 2 | 1 | 2 | 1 | 2 | 1 | 1 | 2 |
| | C1Q4 | 1 | 1 | 1 | 1 | 1 | 1 | 1 | 2 | 1 | 1 |
| **1-after-1** | C1Q5 | 1 | 2 | 3 | 2 | 4 | 1 | 2 | 4 | 3 | 2 |
| | C2Q1 | 1 | 1 | 2 | 1 | 1 | 2 | 2 | 2 | 1 | 1 |
| | C2Q2 | 1 | 1 | 1 | 1 | 3 | 4 | 2 | 1 | 1 | 3 |
| | C2Q3 | 1 | 1 | 2 | 1 | 3 | 1 | 1 | 3 | 2 | 3 |
| | C2Q4 | 1 | 1 | 3 | 2 | 3 | 2 | 3 | 1 | 3 | 1 |
| **No Extra Settings** | C2Q5 | 1 | 1 | 1 | 3 | 3 | 1 | 2 | 3 | 1 | 2 |
| | C3Q1 | 1 | 1 | 1 | 1 | 2 | 2 | 1 | 2 | 1 | 1 |
| | C3Q2 | 1 | 1 | 1 | 2 | 1 | 2 | 1 | 2 | 4 | 2 |
| | C3Q3 | 4 | 1 | 3 | 1 | 2 | 1 | 1 | 3 | 2 | 4 |
| | C3Q4 | 1 | 1 | 2 | 3 | 2 | 3 | 2 | 1 | 5 | 1 |
| **all-in-all** | C3Q5 | 4 | 1 | 2 | 1 | 3 | 1 | 3 | 2 | 1 | 4 |

**Fig. 7.** Difficult Scores Table

| Factor | Mean | Median | Mode |
|---|---|---|---|
| 1-after-1 | 1.9 | 1 | 1 |
| No extra settings | 2.0 | 2 | 1 |
| all-in-all | 2.1 | 1 | 1 |

**Table 2.** Each chatbot's difficulty ratings among all the participants in terms of mean, median, and mode in the first task

Between Table 2 and Table 3, 4 out of 6 median scores are 1 (very easy), and two median scores have been quantified in 2 (easy); one is allocated in the first task's "No extra settings" chatbot, another one is allocated in the second task's "all-in-all" chatbot. The two different median scores occurred in different tasks that happened on different chatbots, along with the mode values for those two special cases are all equivalent to 1 (very easy), implicating the right skewness occurs in two tasks score distributions, resulting in the median being affected and raised up to 2 (easy). Considering participants' experiences, this categorical analysis suggests that chatbots can generate very easy MCQs.

| Factor | Mean | Median | Mode |
|---|---|---|---|
| 1-after-1 | 1.5 | 1 | 1 |
| No extra settings | 1.8 | 1 | 1 |
| all-in-all | 1.9 | 2 | 1 |

**Table 3.** Each chatbot's difficulty ratings among all the participants in terms of mean, median, and mode in the second task

Table 4 has calculated the mean score of each chatbot's capability of generating difficult level's MCQ between two tasks. Even though the average scores that measure the difficulty capability among the chatbots are slightly different, with a maximum marginal variation of 0.3, the results are close to each other as
1.7 and 1.9 can be rounded up to 2 (easy) to interpret the scale rating. The Mean Average Deviation scores are no larger than 0.4, demonstrating each chatbot has a relatively stable capability to generate similar difficulty MCQs for each task. Therefore, all three chatbots have the similar stable capability of generating "easy" difficulty-level MCQs. Based on the previous analysis, there is a contradiction between the difficulty level generated by chatbots and participation perception. Three chatbots can stably generate "easy" level MCQs, while the participation standard view illustrates that questions difficulty-level are "very easy". This has further led to Table 5, which synthesised participants' feedback with respect to their scores to reflect the conflicts. Participants mentioned that "very easy" difficulty levels are too straightforward and basic, consisting of the stable generated "easy" difficulty level scaled from the chatbots. This indicates the chatbot prompts required further refinement about its difficulty-level scales. Additionally, the "very hard" difficulty level is out of the content, which is sup- posed to stay in the content but requires critical thinking and logical judgement from Bloom's taxonomy model as Simkin and Kuechler[11] discussed.

| Factor | Mean |
|---|---|
| 1-after-1 | 1.7 |
| No extra settings | 1.9 |
| all-in-all | 2 |

**Table 4.** The average of difficulty level scores among the chatbots between two tasks

In summary, three chatbots performed similarly in MCQs difficulty generations, with the participants' group common views for three chatbots on difficulty level rated in "1" (very easy), and the average difficulty level scores from each task are close to "2" (easy), which resulted in the conclusion that there's no significant difference among these three discussed adaptive MCQs generating methods to evaluate users' study performance, and they all can measure users' study performance in some extent. However, three chatbots cannot accurately generate stable MCQs with a score of 3 (intermediate level) in expectation and contain the Definition-Perception gap with the defined difficulty level, which requires adjusting the difficulty level to improve.

| | |
|---|---|
| 1 | Intuitive questions and have been asked general definition questions, the correct option is obvious. |
| 2 | Questions are relevant and straightforward with the content of knowledge memorisation, the correct option is quite easy to tell. |
| 3 | Questions require basic knowledge comprehension, and options are able to differentiate, but some terms are quite related. |
| 4 | Questions are mixed up with concepts, require knowledge reflection, options contain distractions that are hard to differentiate. |
| 5 | Questions have covered the content not mentioned, requires in-depth thinking and knowledge application, options contain too many distractions. |

**Table 5.** Participants comment summarisation on difficulty level

## 5 Conclusion and Further Improvements

This study presents MetaCQ, an E-Textbook platform designed to facilitate data privacy learning by integrating adaptive chatbots, OLM-based self-assessment, and metacognitive help-seeking strategies. The platform employs three adaptive feedback mechanisms, evaluated through a user study with the Think-Aloud method. Findings indicate that while the system shows potential, limitations arise from the small participant pool and an imbalance in question difficulty, with most items rated as "very easy." Future work will refine question design by enhancing difficulty differentiation and adjusting adaptive difficulty ranges from
0.1 to 0.3. Moreover, expanding participant involvement for question difficulty annotation and adopting reinforcement learning techniques are suggested to im- prove the generation of higher-difficulty MCQs. Further studies will also apply more robust statistical methods, including hypothesis testing, ANOVA, and t- tests, to validate the adaptive mechanisms' impact across different demographic groups and explore correlations between perceived difficulty, performance, and user engagement.

**Disclosure of Interests.** The authors have no competing interests to declare that are relevant to the content of this article.